# Cr/p–Hg$_{1-x}$Cd$_x$Te (x = 0.28, 1.0) contact resistance and I–V–characteristics in the LTLM configuration

F. Sizov[1], A. Zinovchuk[2], V. Slipokurov[1], Ye. Melezhyk[1*], and Z. Tsybrii[1]

[1] V. Lashkaryov Institute of Semiconductor Physics, National Academy of Sciences of Ukraine, 41 Nauky Avenue, Kyiv 03680, Ukraine
[2] Zhytomyr Ivan Franko State University, 40 Velyka Berdychivska Str., Zhytomyr 10008, Ukraine
* emelezhik@gmail.com

**Abstract**
The specific contact resistance $\rho_c$ of Cr/p–Hg$_{0.72}$Cd$_{0.28}$Te and Cr/p–CdTe/p–Hg$_{0.72}$Cd$_{0.28}$Te heterointerfaces, which were formed by deposition of Cr films at room temperatures on the surfaces of semiconductors, was studied by the Linear Transmission Line Model (LTLM) method. It was observed linear or close to linear behavior of current–voltage (I–V) characteristics at temperatures T = 77 and 300 K are observed. It was assumed that relatively large specific contact resistance $\rho_c$ perhaps is mainly connected with the formation of Cr–oxides and the elements diffusion into Cr layers from the semiconductors under consideration. No intermediate dielectric layer, separating Cr from Hg$_{0.72}$Cd$_{0.28}$Te or CdTe layers was observed. It was accepted that the MIGS (metal-induced gap states) model in the semiconductor can be considered to describe the mechanism of current transport through the Cr/p–CdTe/p–Hg$_{0.72}$Cd$_{0.28}$Te and Cr/p–Hg$_{0.72}$Cd$_{0.28}$Te interfaces (Schottky barriers – SBs). The linear or almost linear I–V–characteristics in the LTLM configuration observed can be explained by taking into account the series resistance of bulk semiconductor substrate and low resistance of large area SBs.

**Keywords:** Cr/p–Hg$_{0.72}$Cd$_{0.28}$Te and Cr/p–CdTe/p–Hg$_{0.72}$Cd$_{0.28}$Te heterointerfaces, specific contact resistance, LTLM method, Schottky barrier, I–V–characteristics.

1. **Introduction**

Today, the cooled quantum infrared (IR) semiconductor detectors (on the base of interband, impurity-band or intersubband transitions) are widely used in various fields of activity: defense, medicine, astronomy, meteorology, remote sensing of the Earth and planets, etc., because they are fast and are more sensitive compared, e.g., with bolometric uncooled detectors. Various semiconductor materials are applied for their design, including mercury–cadmium telluride (MCT – HgCdTe), indium antimonide (InSb), structures based on QWIPS (Quantum Well IR Photodetectors – GaAs/AlGaAs, InGaAs/GaInP), type II superlattices (T2SL), graphene and impurity based detectors, etc. [1 – 3].

Due to its electrical and optical properties, the MCT material was and remains for near future one of the promising materials for SWIR (Short Wavelength IR), MWIR (Medium Wavelength) and LWIR (Long Wavelength) quantum IR detectors [2, 4]. Increasing the operating temperature of the quantum detectors with ultimate performance characteristics is an important task, which seems can be achieved only with MCT detectors [5, 6].

Although this material in a number of electrical and optical properties has advantages over some above–mentioned semiconductor materials, it also requires more thorough study and delicate handling. It concerns, e.g., the weak Hg–Te bonds and the stability of HgCdTe with respect to time and temperature, which can significantly affect the electro–physical characteristics and sensitivity of detectors on its base. Moreover, the resistivity of Hg$_{1-x}$Cd$_x$Te layers changes many orders in dependence of chemical composition "x" that demands careful design of contact formation, especially for SWIR and MWIR detectors.

It is desirable to protect MCT surface by the material or, e.g., some kind of oxides with close lattice (F$\bar{4}$3m cubic lattice) constant α and similar thermal expansion coefficient with similar cubic lattice. Among such materials there can be CdTe (F$\bar{4}$3m cubic lattice), which has α = 6.477 Å, and close thermal expansion coefficient value. MCT lattice constant varies by about 0.3 % over the entire stoichiometric range of $Hg_{1-x}Cd_xTe$ from x = 0 to x = 1 [7]. In several papers there were reported researches on the properties of CdTe/MCT interfaces (see, e.g., [8 – 10].

Such use of CdTe as protective layer is justified by the chemical compatibility, low lattice mismatch, and the reduction of the concentration of parasitic centers of generation, recombination and surface traps of the CdTe/HgCdTe interface. At the same time, the presence of Ohmic contact should minimally affect the change in the electrophysical characteristics of the material itself. As a result, the Ohmic contact formed on the base of the selected contact metallization must satisfy the following requirements: the presence of a linearly symmetrical current-voltage (I–V) characteristics, a low value of the specific contact resistance $\rho_c$, a uniform metal-semiconductor interface.

Various materials were utilized to form contact on HgCdTe like Al, Ag, Cu, Ti, Pd, Pt, Ge, and HgTe. There is a small selection of metals capable of forming an Ohmic contact to p-type HgCdTe. Metals Ag, Cu, Pd, Pt, Sb, Ge demonstrate rectifying behavior, whereas metals Au and Al seems create nearly Ohmic contacts (for Refs. see, e.g., [11]). The purpose of this investigation was to try to reveal the influence of diffusion of elements from MCT epitaxial films or CdTe cap layers into rather thick ($d_{Cr} \approx 460$ nm) adhesion metallic layers deposited at room temperature on the contact resistance of the Cr/p-CdTe and Cr/p–$Hg_{0.28}Cd_{0.72}Te$ structures, which can be applied for designing of MWIR photodetectors. The known Cr high adhesion properties to semiconductors for contact formation to CdTe or MCT surfaces at room temperature, seemed can be an important factor influencing the stability properties of detectors on the base of MCT.

## 2. Technology and Experiment

HgCdTe is a soft material with low energy of Hg–Te bonds [12]. Hence, it should have pronounced chemical interactions at the interface MCT, CdTe and contact metals even at relatively low temperatures due to intermixing and interdiffusion on the CdTe/MCT or metal/MCT interfaces.

### 2.1. p–CdTe/p–$Hg_{0.72}Cd_{0.28}Te$ interface

The CdTe films on $Hg_{1-x}Cd_xTe$ (x ≈ 0.28) monocrystalline (111) B epitaxial layers, grown by Liquid Phase Epitaxy (LPE) method on $Cd_{0.96}Zn_{0.04}Te$ (111) dielectric substrates [13], were formed using the Hot-Wall Epitaxy (HWE) method, which is one of the variations of the thermal gaseous vacuum sputtering methods [14]. In this work mostly closely equilibrium conditions of CdTe films growth takes place. In spite of the relatively low temperatures of the source and the wall and the long deposition time, polycrystalline CdTe films (d ≈ 400 nm grown at temperatures T ≤ 100 °C) with high adhesion to MCT films were obtained. A hot–wall facility mounted in a vacuum chamber enabled to grow films under the residual gases pressure close to 1 × $10^{-6}$ Torr. The same vacuum conditions were at Cr deposition by magnetron sputtering method at substrate at T ≈ 300 K.

To keep the dark current in MCT detectors low and to prevent the degradation of devices during the deposition procedure onto the HgCdTe surface, the low-temperature conditions of growth should be kept to avoid deep depletion of Hg from the HgCdTe surface [15 – 16].

Investigation of the material's quality on the cleaved cross sections of the p–$Hg_{0.72}Cd_{0.28}Te$ grown on $Cd_{0.96}Zn_{0.04}Te$ single crystal dielectric substrate with CdTe passivation (see Fig. 1) shows the lack of impurities in each epitaxial layer (HgCdTe and CdTe). The changes of the Hg–Cd composition in the CdTe (grown at temperatures T ≤ 100 ºC) are due to the diffusion of Hg into the CdTe cap protective layer from the HgCdTe film. The energy-dispersive X-ray (EDX) analysis was carried out to find the elemental and chemical compositions at the surface of grown CdTe/HgCdTe structures. The scanning electron microscope Tescan Mira 3 with energy dispersive spectrometer Oxford Instruments X-max was used. Fig. 1 shows, that Hg is quickly decreasing in CdTe epitaxial layer and is practically absent at the CdTe/vacuum border. The decrease of Hg at 0.5 CdTe layer thickness is ≈4.5 times which is similar to the results of [8] where the Hg decrease in CdTe at 0.5 d thickness is ≈4 times in CdTe protective layer (MBE, T = 260 ºC, t ≈ 130 min).

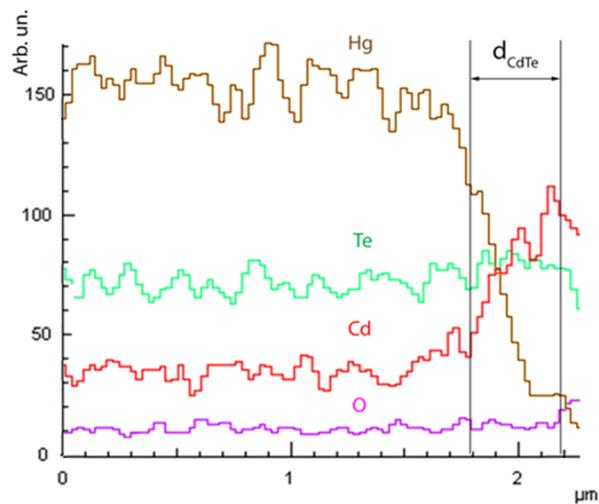

Fig. 1. Distributions of chemical elements in the $Hg_{0.72}Cd_{0.28}Te$/CdTe cross section structure (EDX analysis).

Fig. 2 shows the photograph by a scanning electron microscope (SEM) of the crystal chip of $Hg_{0.72}Cd_{0.28}Te$ film with a passivation CdTe layer. The thickness of the $Hg_{0.72}Cd_{0.28}Te$ epitaxial layer on $Cd_{0.96}Zn_{0.04}Te$ dielectric substrates was $d_{HgCdTe}$ ≈ 15 μm and the thickness of the passivation CdTe layer was d ≈ 400 nm. The CdTe layer was grown by the rate of ≈6 nm/min. As one can see from the SEM photo, the CdTe passivation layer has a typical column–grain structure.

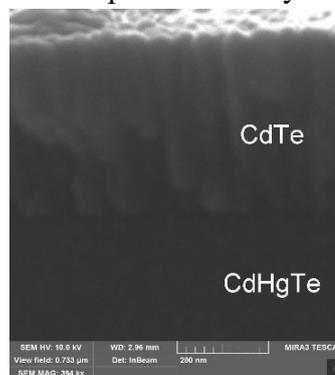

Fig. 2. SEM (Scanning Electron Microscope TescanMira 3 MLU) photo of the cleaved $Hg_{0.72}Cd_{0.28}Te$ epitaxial layer with 400 nm CdTe passivation layer.

The interdiffusion of atoms at the CdTe/HgCdTe interface was also observed in [17], where p–Hg$_{1-x}$Cd$_x$Te (x = 0.225) epitaxial layers were grown by liquid phase epitaxy method and were covered by CdTe dielectric layer with a thickness of 300 nm by electron beam evaporation. The results of secondary ion mass spectrometry show the presence of gradient composition at the CdTe/HgCdTe boundary and additional annealing only increased the magnitude of the gradient. The authors of [17] did not indicate the temperature and other conditions of CdTe growth and annealing of CdTe/HgCdTe structure.

### 2.2. Cr/p–Hg$_{0.72}$Cd$_{0.28}$Te and Cr/p–CdTe/p–Hg$_{0.72}$Cd$_{0.28}$Te contacts

For manufacturing electrical contacts, the room-temperature magnetron sputtering to deposit Cr layers, and thermal evaporation methods to form Au contacts to Cr layers with lift–off lithography were carried out. To study the influence of metallic adhesive layers on the electrical resistance of the contacts to MCT or CdTe films (Fig. 3), here were used the relatively thick (≈460 nm) Cr layers deposited at a growth rate of ≈75 nm/min at a residual gas-pressure of P ≈ 1 × 10$^{-6}$ Torr. Deposition of Cr, as a rule, provides high adhesion characteristics to MCT and CdTe surfaces.

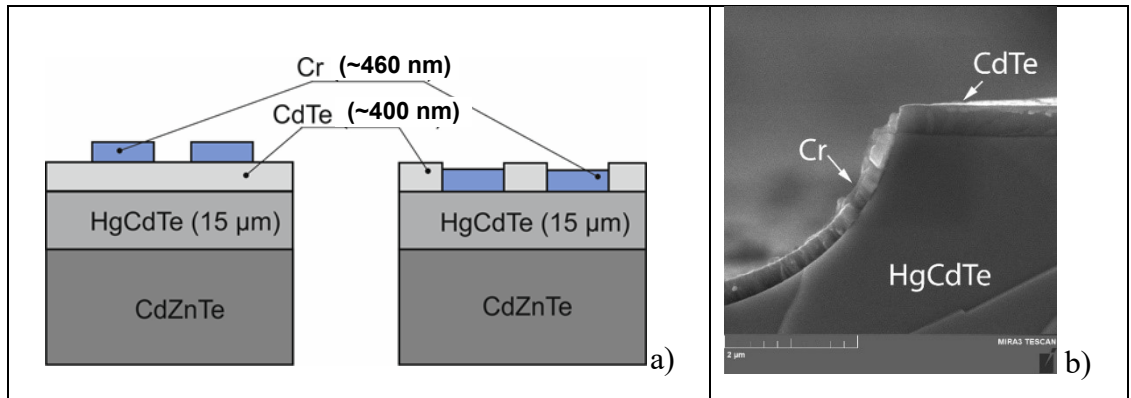

Fig. 3. The schematic cross-section of fabricated structures Cr/p–CdTe/p–Hg$_{0.72}$Cd$_{0.28}$Te /Cd$_{0.96}$Zn$_{0.04}$Te and Cr/p–Hg$_{0.72}$Cd$_{0.28}$Te/Cd$_{0.96}$Zn$_{0.04}$Te (a). SEM photograph of the cleaved part of the sample with Cr/p-HgCdTe interface (TescanMira 3 MLU) (b).

In Fig. 4.a is shown the distribution of elements at MCT/Cr interface structure. It is seen a large amount of oxygen appearing in Cr adhesion layer from the chamber atmosphere together with Te, Hg and Cd components from the MCT layer. In the case of Cr/CdTe heterostructure Hg is absent.

Many crystalline modifications of chromium oxides exist: rutile (CrO$_2$), CrO$_3$, CrO$_4$, corundum (Cr$_2$O$_3$), Cr$_2$O$_5$, and Cr$_5$O$_{12}$. Among these modifications, Cr$_2$O$_3$ is the most stable dielectric oxide–material [18, 19].

To study the elements distribution in rather thick Cr layers on CdTe or MCT it should be taken into account that Cr$_2$O$_3$ has large enthalpy formation energy $\Delta_f H°_{solid}$ = –1134.70 kJ/mol [20]. Cr$_2$O$_3$ is an insulating p-type material with wide band gap of ~ 3.1 – 3.2 eV [19, 21 – 22].

Fig. 4.b demonstrates the areas in which elemental analysis was carried out: CdTe cap layer (Spectrum 1), Cr layer (Spectrum 2) and HgCdTe layer (Spectrum 3). The analysis of the distribution of elements in the 460-nm–thick chromium layer (Fig. 4.b, Table 1, Spectrum 2) confirmed the notable diffusion of Cd, Te, Hg into Cr layer. Approximately the same value of mercury and tellurium atoms diffused from the MCT layer into the Cr layer and smaller amount

of Cd (2.57 at %, 2.96 at % and 0.85 at %, respectively) was registered at the depth of ~230 nm from the HgCdTe surface. These quantities are diminishing quickly to the free surface of Cr (Fig. 4.a). A significant presence of oxygen is observed. Its quantity is quickly growing to the free surface of Cr. The deep penetration of O, Te, Hg into the Cr thick adhesion layer even at technological procedures at room temperature may prevent the conductivity through the thick Cr layer.

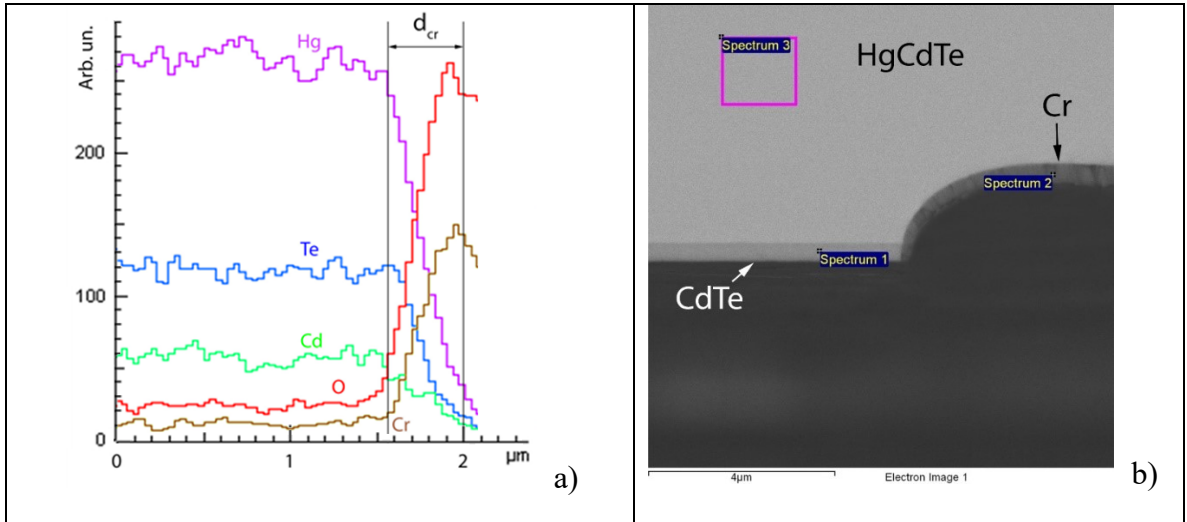

Fig. 4. a) Elements distribution near Cr/p–Hg$_{0.72}$Cd$_{0.28}$Te interface (EDX data, Oxford Instruments X-max); b) SEM photograph with marking of areas in which elemental analysis was carried out (Table 1).

Table 1. Elemental analysis in cross section of Cr/p–Hg$_{0.72}$Cd$_{0.28}$Te structure.

| Element | Spectrum 1, (at ~170 nm from CdTe/ Hg$_{0.72}$Cd$_{0.28}$Te interface) | | Spectrum 2, (at ~230 nm from Cr/ Hg$_{0.72}$Cd$_{0.28}$Te interface) | | Spectrum 3, (Hg$_{0.72}$Cd$_{0.28}$Te film) | |
|---|---|---|---|---|---|---|
| | wt.% | at.% | wt.% | at.% | wt.% | at.% |
| Cd | 41.30 | 45.81 | 2.00 | 0.85 | 10.57 | 14.23 |
| Te | 49.80 | 48.66 | 7.87 | 2.96 | 42.37 | 50.26 |
| Hg | 8.91 | 5.54 | 10.71 | 2.57 | 47.07 | 35.52 |
| Cr | - | - | 69.70 | 64.41 | - | - |
| O | - | - | 9.72 | 29.21 | - | - |
| % | 100.01 % | 100.01 % | 100 % | 100 % | 100.01 % | 100.01 % |

The presence of Cr oxide can increase the resistivity of rather thick Cr layer ($d_{Cr} \approx 460$ nm) and can cause observed (see below) relatively high contact resistance of Cr/CdTe or Cr/Hg$_{0.72}$Cd$_{0.28}$Te structures.

### 2.3. Contact resistance

Two types of contacts were formed, which can be used to design of IR detectors for 3 to 5 μm spectral range: Cr/p–CdTe/p–Hg$_{0.72}$Cd$_{0.28}$Te and Cr/p–Hg$_{0.72}$Cd$_{0.28}$Te. The topology of structures with selected areas for applying contact metallization for further measurements of electro-physical characteristics were chosen for "linear transmission line model" (LTLM) method [11, 23 – 24].

The growth of CdTe epitaxial cap layers with a thickness of $d_{CdTe} \approx 400$ nm was carried out at low temperature conditions T ≤ 100 °C by the HWE. To create electrical contacts to Cr/p–$Hg_{0.72}Cd_{0.28}Te$ and Cr/p–CdTe/p–$Hg_{0.72}Cd_{0.28}Te$ structures, the magnetron sputtering at temperature T ≈ 300K was used for Cr deposition followed by lithographic lifting.

The external contacts to Cr thick films (d ≈ 460 nm) were formed by microsoldering of gold microwires. The value of the specific contact resistance was determined in the initial Cr/p–CdTe/p–$Hg_{0.72}Cd_{0.28}Te$ structure on the $Cd_{0.96}Zn_{0.04}Te$ substrate, then the contact metallization together with the Cr/p–CdTe cap layer was etched in $HBr/Br_2$ methanol solution and the contact metallization layer was re-applied according to the procedure described above. Next, the value of the specific contact resistance of the Cr/p–$Hg_{0.72}Cd_{0.28}Te$ heterostructure was investigated. Determination of the specific contact resistance $\rho_c$ for both structures at temperatures T = 77 and T = 300 K was carried out by measurements of I–V–characteristics between each pair of adjacent Cr contact pads by the LTLM method (Fig. 5). Several sets of test structures with the dimensions of the contact pads with the length L = 40 μm and the width W = 2060 μm were used.

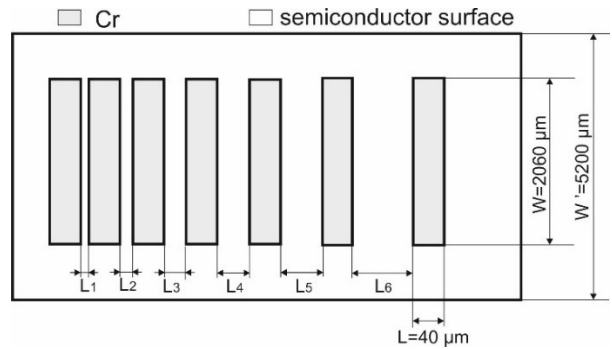

Fig. 5. The scheme of structure for LTLM method with the clearances between Cr contacts in microns: $L_1$ = 400 μm, $L_2$ = 600 μm, $L_3$ = 800 μm, $L_4$ = 1200 μm, $L_5$ = 1600 μm, $L_6$ = 2400 μm.

The measured I–V–characteristics between each of two pairs of the adjacent large area Cr contacts in the LTLM method for both structures are shown in Fig. 6. As can be seen from Fig. 6. a, b, the current–voltage characteristics of the Cr/p–CdTe/p–$Hg_{0.72}Cd_{0.28}Te$ or Cr/p–$Hg_{0.72}Cd_{0.28}Te$ heterostructures have an almost linear-symmetric form at small and large biases at temperatures T = 77 K and T = 300 K.

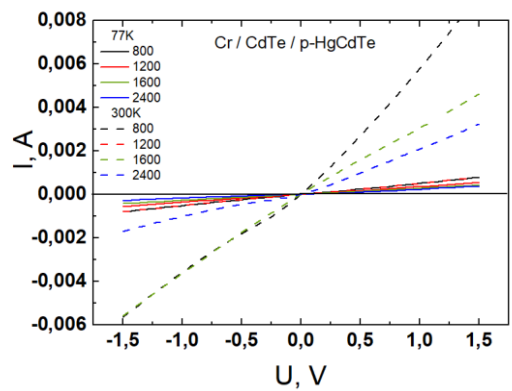

a)

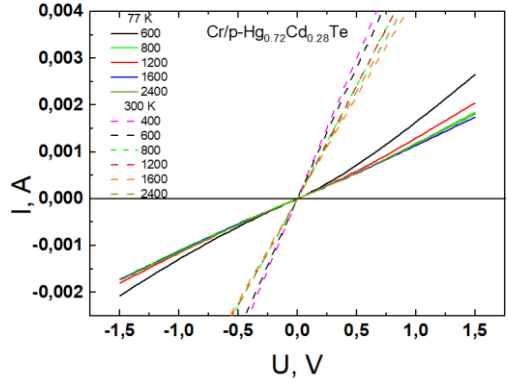
b)

Fig. 6. Current–voltage characteristics of contacts for Cr/p–CdTe/p–Hg$_{0.72}$Cd$_{0.28}$Te (a) and Cr/ p–Hg$_{0.72}$Cd$_{0.28}$Te (b) samples, respectively, at temperatures T = 77 K and T = 300 K. The insets show the distances in μm between the adjacent pairs of contacts.

### 3. Contact resistance methods of measuring

Determination of the specific contact resistance parameter from I–V–curves was carried out by two methods. In the first, the classic LTLM method [23] was applied for a group of linear contacts (method 1):

$$R_i = \frac{R_{sh}L_i}{W} + 2R_c, \tag{1}$$

where $R_i$ is the total resistance between two adjacent Cr contacts, $R_{sh}$ is the surface resistance of a semiconductor, $L_i$ is the distance between two adjacent contact, W is the width of contact areas, $R_c$ is the contact resistance, which also includes the leakage resistance ($R_{sk}$) under the contact. Since $R_{sk} \approx R_{sh}$, the relation (1) can be rewritten in the form (2) as follows

$$R_i = \frac{R_{sh}L_i}{W} + 2\frac{L_T R_{sh}}{W}, \tag{2}$$

where $L_T$ is the transfer length.

Then the point of intersection with the horizontal axis makes it possible to determine $\rho_c$ as follows

$$\rho_c = L_t^2 R_{sh}. \tag{3}$$

The value of $R_{sh}$ can be determined by the angle of inclination of the dependence $R_iW = f(L_i)$. As the length of contact are W, are not equal the width of the sample W' (see Fig. 5), here was used a slightly modified [25] expression to calculate $R_i$.

The second method [25] of determining the value of the specific contact resistance consists in applying the first term of equality (2) to the calculations of the value of the width of the sample W' (Fig. 5) instead of the length of contacts in the case when W = W'.

$$R_i = \frac{R_{sh}L_i}{W'} + 2\frac{\rho_c}{L_t W}, \tag{4}$$

where W' is the sample width.

The dependences $R_iW = f(L_i)$ are shown in (Fig. 7).

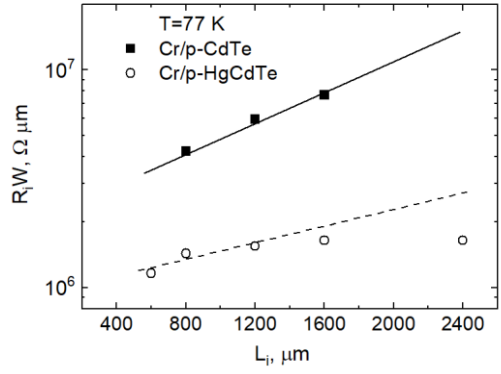

a)

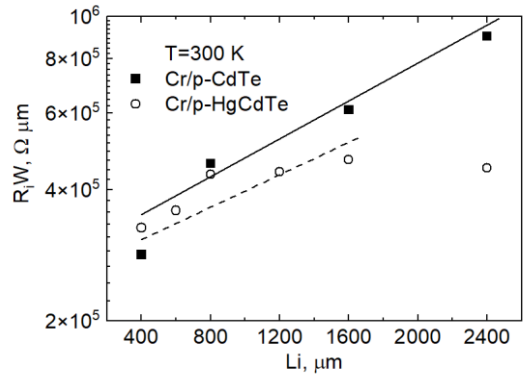

b)

Fig. 7. Typical $R_iW(L_i)$ dependences in the LTLM method for Cr/p–CdTe/p–Hg$_{0.72}$Cd$_{0.28}$Te and Cr/p–Hg$_{0.72}$Cd$_{0.28}$Te structures at temperatures T = 77 K (a) and T = 300 K (b).

The results of calculations of the specific contact resistance for the same sample by the second method are given in Table 2.

Table 2. The specific contact resistance of Cr/p–CdTe/p–Hg$_{0.72}$Cd$_{0.28}$Te and Cr/p–Hg$_{0.72}$Cd$_{0.28}$Te interfaces.

| Parameter | $\rho_c$, 77 K, Ohm·cm$^2$ | $\rho_c$, 300 K, Ohm·cm$^2$ |
|---|---|---|
| Cr/p–CdTe/p–Hg$_{0.72}$Cd$_{0.28}$Te | 0.30 | 0.03 |
| Cr/p–Hg$_{0.72}$Cd$_{0.28}$Te | 0.15 | 0.029 |
| Carrier concentrations, p–CdTe, p–Hg$_{0.72}$Cd$_{0.28}$Te, cm$^{-3}$ | $\approx 3\cdot10^{16}$ | $\approx 3\cdot10^{16}$ |

A difference in the values of specific contact resistance of about 2 times for Cr/p–Hg$_{0.72}$Cd$_{0.28}$Te interface and Cr/p–CdTe/p–Hg$_{0.72}$Cd$_{0.28}$Te one, can be related to the difference in diffusion of Hg$_{0.72}$Cd$_{0.28}$Te or CdTe components into the Cr adhesion layer during the growth of the Cr layer even at room temperature. In the case of the presence of the CdTe cap layer on the Hg$_{0.72}$Cd$_{0.28}$Te epitaxial film, the Hg component almost does not penetrate through the relatively thick ($d_{CdTe} \approx 400$ nm) CdTe layer (Fig. 1) into the Cr ($d_{Cr} \approx 460$ nm) adhesive layers of Cr/CdTe/Hg$_{0.72}$Cd$_{0.28}$Te structure. Therefore, it seems that the main contribution to the specific contact resistance is connected with the Hg diffusion into the Cr layer and Cr oxides appearing in the Cr film during its growth procedure.

## 4. Model and calculations

The results of experimental measurements at T = 77 and 300 K are compared with the numerical simulations of carrier transport through Cr/p–CdTe/p–Hg$_{0.72}$Cd$_{0.28}$Te and Cr/p–Hg$_{0.72}$Cd$_{0.28}$Te MS (Metal-Semiconductor) interfaces. The comparison was made for different mechanisms of charge carrier transport. Electrical properties of the MS interfaces are discussed from the point of view of general considerations about the formation of Schottky barrier height (SBH).

In the classical Schottky–Mott model, the contact of wide–band-gap doped p–CdTe ($E_g \approx 1.5$ eV) with a metal should lead to nonlinear I–V–characteristics due to the high work function energy of CdTe (see Table 3). For Hg$_{1-x}$Cd$_x$Te across x = 0 to x = 1, from the Table 3 it seems that the work function is within $\Phi_S \approx 5.6 - 5.9$ eV. Therefore, in the calculating estimations below for chemical composition x = 0.3, it was taken $\Phi_S \approx 5.6$ eV. This value does not match the work function of Cr ($\Phi_M \approx 4.5$ eV) or any kind of metal perhaps except of Pt (depending on cleanliness of the surface, poly– or single–crystal and crystallographic orientation).

Table 3. Work functions of some metals and HgCdTe summarized in [26].

| Metal, MCT | Au | Cr | In | Mo | Ni | Sn | Pt | Ti | CdTe | HgTe | Hg$_{0.8}$Cd$_{0.2}$Te |
|---|---|---|---|---|---|---|---|---|---|---|---|
| Work function, eV | 5.31–5.47 | 4.5 | 4.09 | 4.50–4.95 | 5.22–5.35 | 4.42 | 5.12–5.93 | 4.33 | 5.7–5.9 | ≈5.9 | 5.6 |

Therefore, all contacts to p–CdTe or p–Hg$_{0.72}$Cd$_{0.28}$Te should be rectifying according to the classical Schottky–Mott model. To explain the Ohmic (or slightly rectifying) I–V–behavior (Fig. 6) obtained within the LTLM experiments, the series resistance of grown semiconductor layers on Cd$_{0.96}$Zn$_{0.04}$Te dielectric substrates, should also be taken into account and one should assume the presence of the Fermi level pinning at the MS interface.

Many models for Schottky barrier physics have been developed (for Refs. see, e.g., [27, 28]). Despite the long research history, the basic mechanisms of Fermi level pinning phenomenon at MS interfaces have not been well established yet. The Surface State Model [29, 30], the Metal–Induced Gap States (MIGS) model [31, 32]; the Unified Defect model [33], the Disorder–Induced Gap States (DIGS) model [34] are known as the most applicable Fermi level pinning models for the MS interface investigations. In this study, we follow the MIGS model, in which decaying into the semiconductor at small distances metal wave functions induce the metal-induced gap states. This model was used to account for the pinning effect at Cr/p–CdTe and Cr/p–HgCdTe interfaces as there were not revealed any dielectric layers at the MS interface (Fig. 4).

Including the pinning effect, the SBH (Schottky Barrier Height) is given for a p-type semiconductor by [30]

$$\Phi_{bp} = \gamma(\Phi_M - \chi_S) + (1-\gamma)(E_g - E_{cnl}) - E_g, \qquad (5)$$

where $E_g$ is the semiconductor band-gap ($E_{g\text{CdTe}} = 1.58$ and 1.46 eV for 77 and 300 K, respectively; $E_{g\text{Hg0.72Cd0.28Te}} = 0.24$ and 0.29 eV for T = 77 and T = 300 K, respectively); $E_{cnl}$ is the location of the Charge Neutrality Level (CNL) with respect to the top of semiconductor valence band; $\gamma$ is the parameter describing the degree of Fermi level pinning. Within the framework of the MIGS model, $\gamma$ can be estimated as

$$\gamma = \left(1 - \frac{e^2 D_{gs} \delta}{\varepsilon \varepsilon_0}\right)^{-1}, \qquad (6)$$

where $e$ is the elementary charge; $\varepsilon_0$ is the electric constant; $\varepsilon$ is the semiconductor dielectric constant; $\delta$ and $D_{gs}$ are the penetration depth and the surface density of metal-induced gap states per energy (in cm$^{-2}$·eV$^{-1}$ units) which is supposed to be roughly constant near the CNL.

Carrier transport through the MS interface is described with the classical Thermionic Emission (TE), Tunneling Field Emission (FE), and Thermionic Field Emission (TFE) mechanisms. Because of the fact that the tunneling mechanism is extremely sensitive to the Schottky barrier profile [35], the analytical expressions for tunneling current derived within numerous simplifications can lead to questionable conclusions.

To obtain more reliable results, numerical calculations of the carrier transport should be performed. In this work we use a general expression for the current in metal/*p–type semiconductor* structures which includes both thermionic as well as tunneling mechanism of carrier transport

$$J = \frac{4\pi e m^*}{h^3} \int_{-\infty}^{0} \tau(E_x) \int_{-\infty}^{0} \left( f_S(F_S - E_x - E_{yz}) - f_M(E_x + E_{yz} - F_M) \right) dE_{yz} dE_x, \quad (7)$$

where $h$ is Planck's constant; $E_x$ and $E_{yz}$ are the normal and parallel components of particle energy with respect to the barrier orientation; $m^*$ is the effective mass of tunneling particle; $f_S$ ($f_M$) and $F_S$ ($F_M$) are the distribution function and Fermi level position in p-type semiconductor and metal, respectively; $\tau(E_x)$ is the transmission coefficient. In the Wentzel–Kramers–Brillouin approximation $\tau(E_x)$ is equal to

$$\tau(E_x) = \exp\left( -\frac{4\pi\sqrt{2m^*}}{h} \int_{x_1}^{x_2} \sqrt{E_x - \varphi(x)} dx \right), \quad (8)$$

where $\varphi(x)$ is the potential function describing the Schottky barrier profile; $x_1$ and $x_2$ are the classical turning points. We note that $E_x$ and $E_{yz}$ in Eq. (7) and (8) are negative (measured from the top of semiconductor valence band).

Having in mind that in HgCdTe semiconductor the heavy hole mass is much higher that the light hole mass, it was neglect the heavy hole transmission coefficient assuming only light hole tunneling ($m^* = 0.014m_0$ for Hg$_{0.72}$Cd$_{0.28}$Te and $m^* = 0.11m_0$ for CdTe). The condition of $\tau(E_x) = 1$ in Eq. (7) represents the thermionic mechanism for the holes, which have the energy $E_x$ high enough to overcome the Schottky barrier. In thermal equilibrium, the Fermi level positions in both the semiconductor and metal ($F_M = F_S$) are calculated from the experimental hole concentrations at T = 77 and T = 300 K (Table 2). When a voltage $V$ is applied, the Fermi level in the semiconductor shifts upward (reverse bias $V < 0$) or downward (forward bias $V > 0$) with respect to the metal level following the relation $F_M - F_S = eV$.

The last input parameter, which have to be discussed to complete the numerical model, is the potential function $\varphi(x)$ in Eq. (8). Following the Thin Surface Barrier (TSB) model originally proposed in [36] and refined in [35], it was assumed that MS gap states produce a high density of unintentional donors or acceptors (as in this case) near the MS interface, which flatten or sharpen a conventional space charge region of SB profile in the depletion approximation.

Within the MIGS penetration depth $x < \delta$, the Schottky barrier profile is defined by the surface potential. The form of the surface potential is a sharp parabola parameterized with two parameters: the surface barrier height $E_s$ and location of potential minimum $d_s$. In the case of $\delta < x < d$ (where $d$ is the space charge region thickness), the Schottky barrier profile is calculated in the depletion approximation using bulk acceptor doping density ($N_a \approx 3 \cdot 10^{16}$ cm$^{-3}$) and bulk valence band bending $V_{bb}$. Fig. 8. schematically shows a resulting view of the potential function $\varphi(x)$ and the meaning of the defining parameters. We put the value of the surface barrier height $E_s$ to be equal to $\Phi_M - \chi_S - E_g$ (according to the Schottky–Mott rule). The bulk valence band bending is $V_{bb} = \Phi_{bp} - F_S$. The location of potential minimum $d_s$ serves as a parameter to ensure the potential function continuity at $x = \delta$.

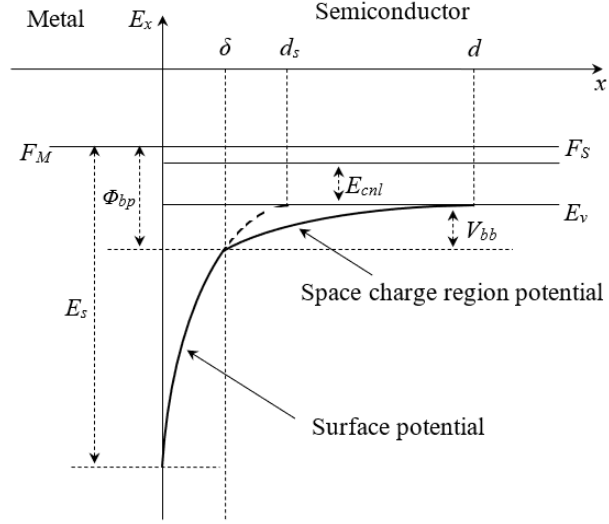

Fig. 8. Schematic representation of the Schottky barrier profile in the MIGS model. Note that the charge neutrality level $E_{cnl}$, bulk valence band bending $V_{bb}$ are referenced to the to the valence band $E_v$ while the SBH $\Phi_{bp}$ and surface barrier height $E_s$ are referenced to the Fermi level position $F_M$. The dimension parameters $\delta$, $d_s$ and $d$ are the MIGS penetration depth, location of the surface potential minimum and the space charge region thickness. Numerical values for the parameters depicted in the scheme are obtained (at a given temperature and applied voltage $V$) from Eq. 5 and Eq. 6 using the data from Table 4.

Table 4. List of model fitting parameters.

|  | Cr/p–CdTe | Cr/p–Hg$_{0.72}$Cd$_{0.28}$Te |
|---|---|---|
| $E_{cnl}$ (eV) | $0.07E_g$ | $0.27E_g$ |
| $\delta$ (nm) | 1.0 | 5.0 |
| $D_{gs}$ (cm$^{-2}$ V$^{-1}$) | $4 \cdot 10^{15}$ | $5 \cdot 10^{14}$ |

To fit I–V–characteristics at forward biases, it was incorporated the series resistance $R_s$ into the model using an approximate relation $R_s = e p \mu_p / L$, where $p$ is the experimental hole concentrations (Table 2); $\mu_p$ is the hole mobility; $L$ is the distance between adjacent contact pads in Fig. 5. Next typical values are used for the hole mobility: $\mu_p$(CdTe) = 300 and 70 cm$^2\cdot$V$^{-1}\cdot$s$^{-1}$ for T = 77 and T = 300 K; $\mu_p$(Hg$_{0.72}$Cd$_{0.28}$Te) = 400 and 40 cm$^2\cdot$V$^{-1}\cdot$s$^{-1}$ for T = 77 and T = 300 K.

Summarizing the theoretical description, the model has three parameters ($E_{cnl}$, $\delta$ and $D_{gs}$) that can not be evaluated in a straightforward way. We treat these physical quantities as fitting parameters to match the experimental data. In this work, we use a two–objective self–adjusted fitting procedure: the first objective is the experimental values of the specific contact resistance (Table 2); the second objective is the linearity of I–V–characteristics.

The results of the fitting procedure are presented in Table 4. The obtained surface densities of MIGS are slightly higher than those reported in [30] for Si, GaAs and GaP metal–semiconductor contacts. The location of the CNL with respect to the top of semiconductor valence band for both p–CdTe and p–Hg$_{0.72}$Cd$_{0.28}$Te is very close (for example at T = 77 K, $E_{cnl}$ = 0.11 and 0.08 eV). However, a large difference is obtained in the MIGS penetration depths. The calculations suggest that the MIGS penetration depth at Cr/p–CdTe interface is much smaller as compared to Cr/p–Hg$_{0.72}$Cd$_{0.28}$Te structure.

Using the values from Table 4 and Eq. (7), we have performed I–V–modeling of the MS interfaces. The results show that the model better reproduces the electrical properties of Cr/p–Hg$_{0.72}$Cd$_{0.28}$Te interface. The calculated values for the specific contact resistance for Cr/p–Hg$_{0.72}$Cd$_{0.28}$Te are 0.2 (T = 77 K) and 0.03 (T = 300 K) Ohm·cm$^2$, which are close to the experimental values. However, for Cr/p–CdTe/p–Hg$_{0.72}$Cd$_{0.28}$Te calculated values are 0.8 (T = 77 K) and 0.009 (T = 300 K) Ohm·cm$^2$, which are in poorer agreement with the experimental data.

In Figs. 9 are shown the estimated Schottky barrier resistance values versus voltage dependences for large area (A = 40 μm × 2060 μm = 8.24·10$^{-4}$ cm$^{-2}$) Cr/p–Hg$_{0.72}$Cd$_{0.28}$Te interfaces (in the LTLM configuration, Fig. 5) with two different contacts at distances L$_1$ = 400 μm and L$_2$ = 600 μm and at T = 300 K and T = 77 K, respectively. One can see that in both cases the Schottky barrier resistances are smaller than the experimentally determined series resistances data between the two adjacent barrier contacts (R$_1$ = 168 Ω for L$_1$ = 400 μm, T = 300 K, R$_2$ = 810 Ω for L$_2$ = 600 μm, T = 77 K).

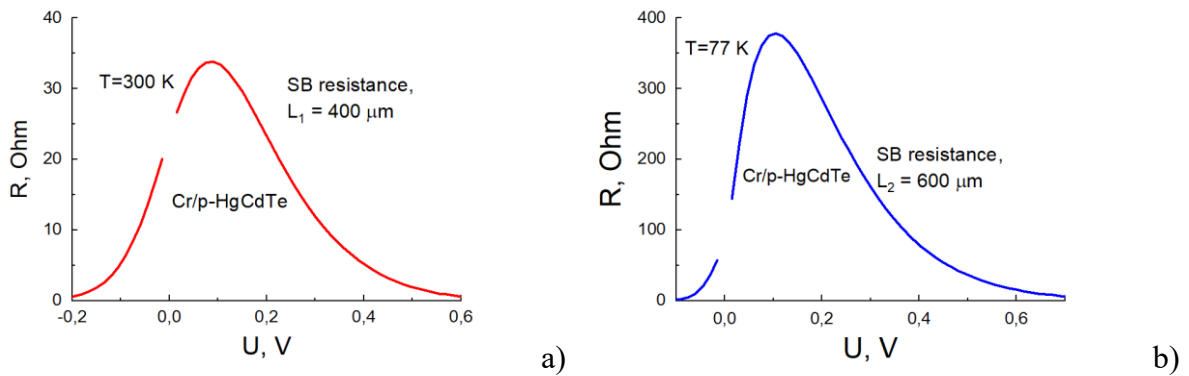

Fig. 9. Calculated SB resistances values versus voltage between a pair of Cr/p–Hg$_{0.72}$Cd$_{0.28}$Te contacts in the LTLM configuration at T = 300 K (a) and T = 77 K (b).

Fig. 10 demonstrates the simulated I–V characteristics of Cr/p–Hg$_{0.72}$Cd$_{0.28}$Te MS interfaces at T = 77 and T = 300 K according to the parameters listed in Table 4. The calculated I–V–curves demonstrate practically linear dependences at T = 300 K. They are slightly deviated from linearity at T = 77 K. In Fig. 10 are shown only the parts of I–V–curves at forward biases as at reverse biases the I–V–curves are symmetrical, because of two adjacent Schottky barriers are symmetrically disposed one to each other. When the forward bias is applied to one of the Scottky barriers, the counterpart one will be reverse biased, and vice versa. The both curves are in a reasonable correspondence with the experimental data. The larger the Schottky barrier resistance is, the larger deviation from the linear experimental I–V–dependences will be observed.

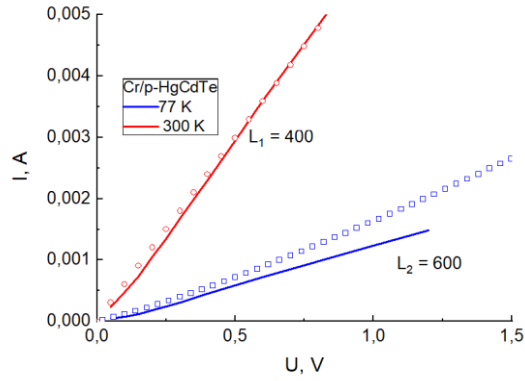

Fig. 10. Calculated I–V–dependences between a pair of adjacent Schottky contacts in the LTLM configuration (solid lines) and experimental data (symbols).

The results obtained show that the linear or almost linear I–V–characteristics measured in the LTLM configuration can't prove the Ohmic contacts behavior of large area Schottky contacts in the LTLM method. The small nonlinearity in I–V–dependences can be seen only at small biases around U = 0, when the resistance of large area SDs is comparable with the series resistance of the semiconductor layers.

## 5. Conclusions

The Cr/p–$Hg_{0.72}Cd_{0.28}Te$ and Cr/p–CdTe/p–$Hg_{0.72}Cd_{0.28}Te$ heterostructures were formed at residual vacuum pressure P ≈1 × $10^{-6}$ Torr by magnetron sputtering at temperature T = 300 K. The specific contact resistance $\rho_c$ of Cr/p–semiconductor heterointerfaces was studied by the LTLM method. It was admitted that the relatively large specific contact resistance $\rho_c$ perhaps is mainly connected with the formation of Cr–oxides and the elements diffusion from the semiconductors under consideration into Cr layers. To explain the linear, or close to linear behavior of I–V–characteristics observed at temperatures T = 77 and 300 K, the series resistance of Schottky barriers (large area Cr layers on p-type $Hg_{1-x}Cd_xTe$ semiconductor layers in LTLM configuration) and the resistance of semiconductor films were considered. The MIGS model was applied to presume the mechanism of current transport through the Cr/p–CdTe/p–$Hg_{0.72}Cd_{0.28}Te$ and Cr/p–$Hg_{0.72}Cd_{0.28}Te$ interfaces. Accounting of the relatively large series resistance of bulk semiconductor films and large area resistance of Schottky barriers, explain the linear or almost linear I–V–characteristics in the LTLM configuration for these structures. Therefore, the linearity of I–V–characteristics in the LTLM configuration can not be used to prove the Ohmic behavior of metal contacts to p–CdTe/p–$Hg_{0.72}Cd_{0.28}Te$ or p–$Hg_{0.72}Cd_{0.28}Te$ layers.

**Acknowledgments**
The authors are thankful to K. Svezhentsova, M. Vuichyk and N. Dmytruk for carrying out some technological operations.
Funding:
This work was partly supported by the Volkswagen Foundation Partnerships-Cooperation Project "Terahertz optoelectronics in novel low-dimensional narrow-gap semiconductor nanostructures" (project number 97738) and the NAS of Ukraine, project No. III-10-24.
Conflict of Interest:
The authors have no conflicts to disclose.

Author Contributions:
F. Sizov - Writing – original draft (lead); Conceptualization (lead); Validation (equal). A. Zinovchuk - Investigation (lead); Software (lead); Writing – review & editing (equal). V. Slipokurov - Methodology (lead); Data curation (equal); Formal analysis (equal). Ye. Melezhyk - Validation (supporting); Writing – review & editing (equal). Z. Tsybrii - Formal analysis (equal); Data curation (equal).

Data availability:
The data that support the findings of this study are available within the article.